\documentclass[twocolumn, prl, notitlepage]{revtex4-1}
\usepackage{amssymb}

\usepackage{graphicx}
\usepackage{dcolumn}
\usepackage{bm}
\usepackage[breaklinks=true,colorlinks=true,linkcolor=blue,urlcolor=blue,citecolor=blue]{hyperref}
\usepackage{physics}

\begin{document}

\title{Fast Generation of GHZ-like States Using Collective-Spin XYZ Model}
\author{Xuanchen Zhang$^{1}$}
\thanks{These authors contributed equally.}
\author{Zhiyao Hu$^{2,1}$}
\thanks{These authors contributed equally.}
\author{Yong-Chun Liu$^{1,3}$}
\email{ycliu@tsinghua.edu.cn}
\affiliation{$^{1}$State Key Laboratory of Low-Dimensional Quantum Physics, Department of
Physics, Tsinghua University, Beijing 100084, China}
\affiliation{$^{2}$School of Physics, Xi'an Jiaotong University, Xi'an
710049, China}
\affiliation{$^{3}$Frontier Science Center for Quantum
Information, Beijing 100084, China}
\date{\today }

\begin{abstract}
    The Greenberger-Horne-Zeilinger (GHZ) state is a key resource for quantum
    information processing and quantum metrology. The atomic GHZ state can be
    generated by one-axis twisting (OAT) interaction $H_{\mathrm{OAT}}=\chi
    J_{z}^{2}$ with $\chi $ the interaction strength, but it requires a long
    evolution time $\chi t=\pi /2$ and is thus seriously influenced by
    decoherence and losses. Here we propose a three-body collective-spin XYZ
    model which creates a GHZ-like state in a very short timescale $\chi t\sim
    \ln {N}/N$ for $N$ particles. We show that this model can be effectively
    produced by applying Floquet driving to an original OAT Hamiltonian.
    Compared with the ideal GHZ state, the GHZ-like state generated using our
    model can maintain similar metrological properties reaching the
    Heisenberg-limited scaling, and it shows better robustness to decoherence
    and particle losses. This work opens the avenue for generating GHZ-like
    states with a large particle number, which holds great potential for the study
    of macroscopic quantum effects and for applications in quantum metrology and
    quantum information.
    \end{abstract}
    
    \maketitle

    
    
    The Schr\"{o}dinger cat state \cite{Schrodinger1935}, a class
    of quantum superpositions of macroscopically distinct states, is crucial in
    investigating large-scale quantum effects \cite{Frowis2018}, understanding
    the quantum-to-classical transition \cite{Brune1996} and numerous
    applications in quantum information science \cite%
    {Zhao2004,Knill2005,Mirrahimi2014} and quantum metrology \cite%
    {Munro2002,Giovannetti2004,Degen2017,Pezze2018}. Many efforts have been made
    to demonstrate such states in diverse systems, including trapped ions \cite%
    {Wineland2013,Monroe1996,Sackett2000,Monz2011,Lo2015}, photonic systems \cite%
    {Alexei2006,Deleglise2008,Gao2010,Huang2015,Chen2016,Sychev2017,Hacker2019,Zhiling2022}%
    , superconducting circuits \cite{Friedman2000,DiCarlo2010,Song2017,Song2019}%
    , Rydberg atoms \cite{Omran2019}, solid states \cite{Bild2023}, and so on.
    Some typical types of the Schr\"{o}dinger cat state are especially attractive,
    such as the entangled coherent state of photons \cite%
    {Joo2011,Sanders2012,Zhang2013} and the multiparticle maximally entangled
    state, which is often referred to as the Greenberger-Horne-Zeilinger (GHZ) state
    \cite{Greenberger1990} or NOON state \cite{Lee2002}. In addition to its
    fundamental significance of demonstrating quantum nonlocality \cite%
    {Greenberger1990,Mermin1990}, the GHZ state can also be applied in quantum
    metrology to achieve Heisenberg-limited measurement
    precision \cite{Bollinger1996,Leibfried2004}, which is even better than the
    optimal performance of squeezed spin states
    \cite{Kitagawa1993,Wineland1994,Sorensen2001prl,Jin2009,Liu2011,Ma2011,Shen2013,Zhang2014,Pezze2018,Chen2019,Huang2021,Hu2023,Huang2023,Bornet2023,Law2001,Helmut2014,Muessel2015,Sorelli2019}.
    
    To generate the GHZ state, entangling gates between quantum bits can be used \cite{Song2019,Choi2014,Barends2014,Kaufmann2017,Lu2019,Wei2020}, but the scale is very small.
    A Large-scale GHZ state can be generated using the collective-spin one-axis twisting (OAT) model $H_{\mathrm{OAT}}=\chi
    J_{z}^{2}$ \cite{Kitagawa1993}, but the required time $\chi t=\pi /2$ is quite long, setting obstacles for
    realizations due to decoherence and particle losses \cite{Sorensen2001,Pezze2009,Agarwal1997,Molmer1999,Zheng2001,Leibfried2005,Chalopin2018}. Measurement and postselection can be used to shorten the evolution time at the expense of the probability of success \cite{Alexander2020}.
    Further studies show that by adding a turning term $\propto J_{x}$ to a OAT
    Hamiltonian, a many-particle entangled state that resembles GHZ state can be
    obtained in a short timescale \cite{Micheli2003,Huang2006}, but it differs from the ideal GHZ state in
    its wider probability distribution on the Fock-space basis.
    
    Here we present a three-body collective-spin interaction model $H_{%
    \mathrm{XYZ}}\propto J_{x}J_{y}J_{z}+J_{z}J_{y}J_{x}$ (which we name the XYZ
    model) to rapidly create a GHZ-like state possessing a Heisenberg-limited
    metrological property and even being much more robust than ideal GHZ state.
    Using the XYZ model, the time to obtain the GHZ-like state scales as $\chi t\sim
    \ln {N}/N$, representing a great shortcut compared with the OAT model,
    especially for a large number of particles $N$. Considering practical
    realizations, we propose a Floquet driving scheme to effectively produce the XYZ
    model from an original OAT Hamiltonian. The metrological property of the
    obtained GHZ-like state shows a Heisenberg-limited
    precision in parity measurements. The degradation of the GHZ-like state due to decoherence and
    particle losses is much slighter than that of the ideal GHZ state, enabling its
    application in noisy environments.
    
    We introduce a collective-spin cubic interaction model for $N$ spin-1/2
    particles, with the Hamiltonian given by
    \begin{equation}
    H_{\mathrm{XYZ}}=\frac{2\chi _{\mathrm{XYZ}}}{N}%
    (J_{x}J_{y}J_{z}+J_{z}J_{y}J_{x}),  \label{XYZ}
    \end{equation}%
    where $J_{\mu }=\sum_{k=1}^{N}\sigma _{\mu }^{(k)}/2$ $(\mu =x,y,z)$ in
    terms of Pauli matrices $\sigma _{\mu }^{(k)}$ denotes the component of the
    collective spin operator $\mathbf{J}$ with total spin $j=N/2$, and $\chi _{%
    \mathrm{XYZ}}$ is the interaction strength. Since the Hamiltonian is
    symmetric under cyclic permutations of $J_{x},J_{y}$ and $J_{z}$, we refer
    to it as the XYZ model hereafter.
    
    \begin{figure*}[tbp]
    \includegraphics[width=2\columnwidth]{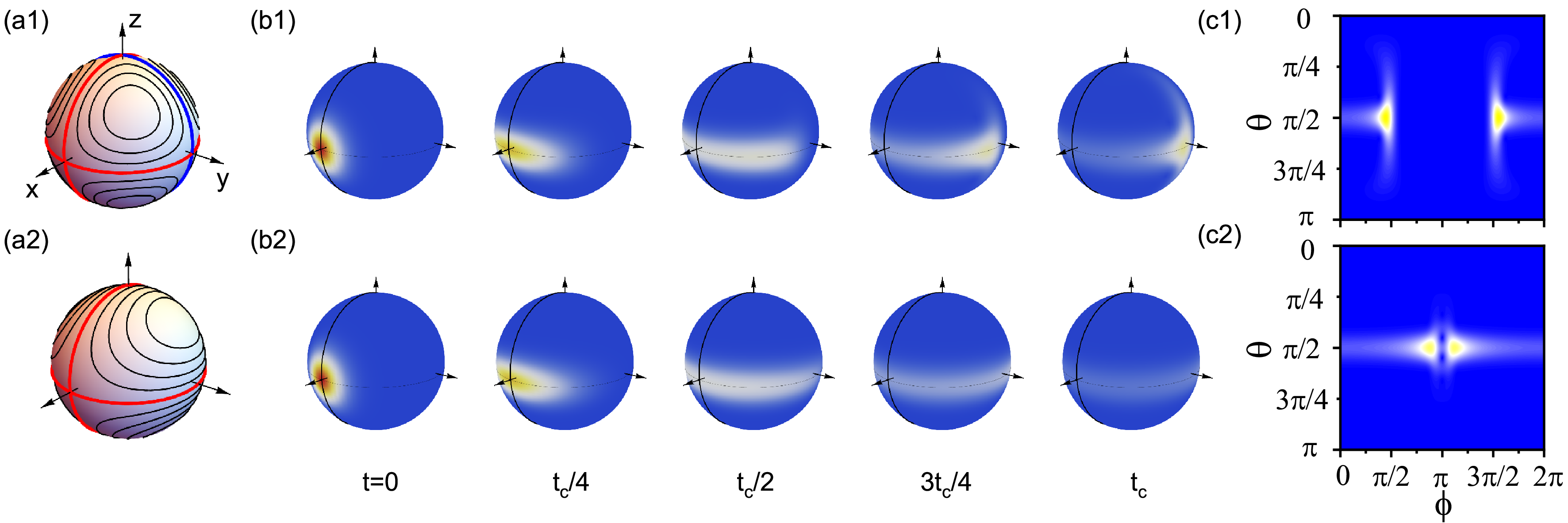}
    \caption{Dynamics of the XYZ model and the TAT model, with $\protect\chi _{%
    \mathrm{XYZ}}=\protect\chi _{\mathrm{TAT}}$. (a) Classical trajectories of
    (a1) the XYZ model and (a2) the TAT model on the Bloch sphere.
    The red trajectories are separatrixes shared by the XYZ model and the TAT
    model, while the blue one is one that only exists in the XYZ model. (b)
    Evolution of the quantum states of (b1) the XYZ model and (b2) the TAT
    model, represented by the Husimi $\mathcal{Q}$ function on the Bloch spheres. The evolution time is labeled with $t_{c}=\ln {N}/(N\protect%
    \chi _{\mathrm{XYZ}})$. (c) Husimi $\mathcal{Q}$ function of (c1) the XYZ
    model and (c2) the TAT model at $t=t_{c}$, represented with polar
    coordinates $(\protect\theta ,\protect\phi )$.}
    \label{fig:1}
    \end{figure*}
    
    As plotted in Fig.~\ref{fig:1}, starting from an initial coherent spin state $%
    \ket{x}$ satisfying $J_{x}\ket{x}=j\ket{x}$, we find that the evolution
    under $H_{\mathrm{XYZ}}$ can stretch the state along the equator and\ split
    it into a GHZ-like state, a coherent superposition of two states
    concentrating near $(0,1,0)$ and $(0,-1,0)$ on the Bloch sphere,
    respectively. For a short timescale before reaching the GHZ-like state, the
    state evolution is similar to the well-known two-axis twisting (TAT) model
    \cite{Kitagawa1993,Liu2011} $H_{\mathrm{TAT}}=\chi _{\mathrm{TAT}}(J_{\pi
    /4,\pi /2}^{2}-J_{\pi /4,-\pi /2}^{2})=\chi _{\mathrm{TAT}%
    }(J_{y}J_{z}+J_{z}J_{y})$, where $J_{\pi /4,\pm \pi /2}=(J_{y}\pm J_{z})/%
    \sqrt{2}$, corresponding to two twisting axes along $\pm 45^{\circ }$
    directions in the $y$-$z$ plane. By comparing the semiclassical
    trajectories \cite{Supp} as shown in Fig.~\ref{fig:1}(a), we can find that
    the XYZ and TAT models behave similarly near the $x$ poles $(\pm 1,0,0)$ on
    the Bloch sphere, with the same separatrixes described by the great circles
    in the $x$-$y$ and $x$-$z$ planes (red circles); but they behave differently
    near the $y$ poles $(0,\pm 1,0)$, where the XYZ model has an additional
    separatrix described by the great circles in the $y$-$z$ planes (blue
    circles), which leads to two additional fixed points at $(0,\pm 1,0)$.
    Therefore, as shown in Fig.~\ref{fig:1}(b), for short $t$ when the state is
    distributed near the pole $(1,0,0)$, the evolution under the XYZ model
    resembles the TAT model with the stretch of the state along the equator, but
    the evolutions diverge for longer time. For the TAT model, the state
    evolution will cross over the points $(0,\pm 1,0)$ and finally converge near
    the point $(-1,0,0)$ ($\phi =\pi $), while for the XYZ model, the state
    distribution can be split into two symmetric parts near the fixed points $%
    (0,\pm 1,0)$, as it sets stoppages at these fixed points. The comparison is
    shown more clearly in Fig.~\ref{fig:1}(c), where the state in the XYZ model
    is located near $(\theta ,\phi )=(\pi /2,\pi /2)$ and $(\pi /2,3\pi /2)$ in
    the polar coordinates, with the separation reaching the maximum.
    
    To obtain the critical evolution time for generating the GHZ-like state in
    the XYZ model, we can use the semiclassical treatment by considering the
    evolution of the point at the edge of the uncertainty patch \cite%
    {Munoz2023,Supp}, which approximately yields
    \begin{equation}
    t_{c}\simeq \frac{\ln {N}}{\chi _{\mathrm{XYZ}}N}.
    \end{equation}
    It reveals that the required time can be very short for large particle
    number $N$, far less than that of the OAT dynamics, which is of crucial importance for overcoming decoherence and
    losses.
    
    \begin{figure}[tbp]
    \includegraphics[width=\columnwidth]{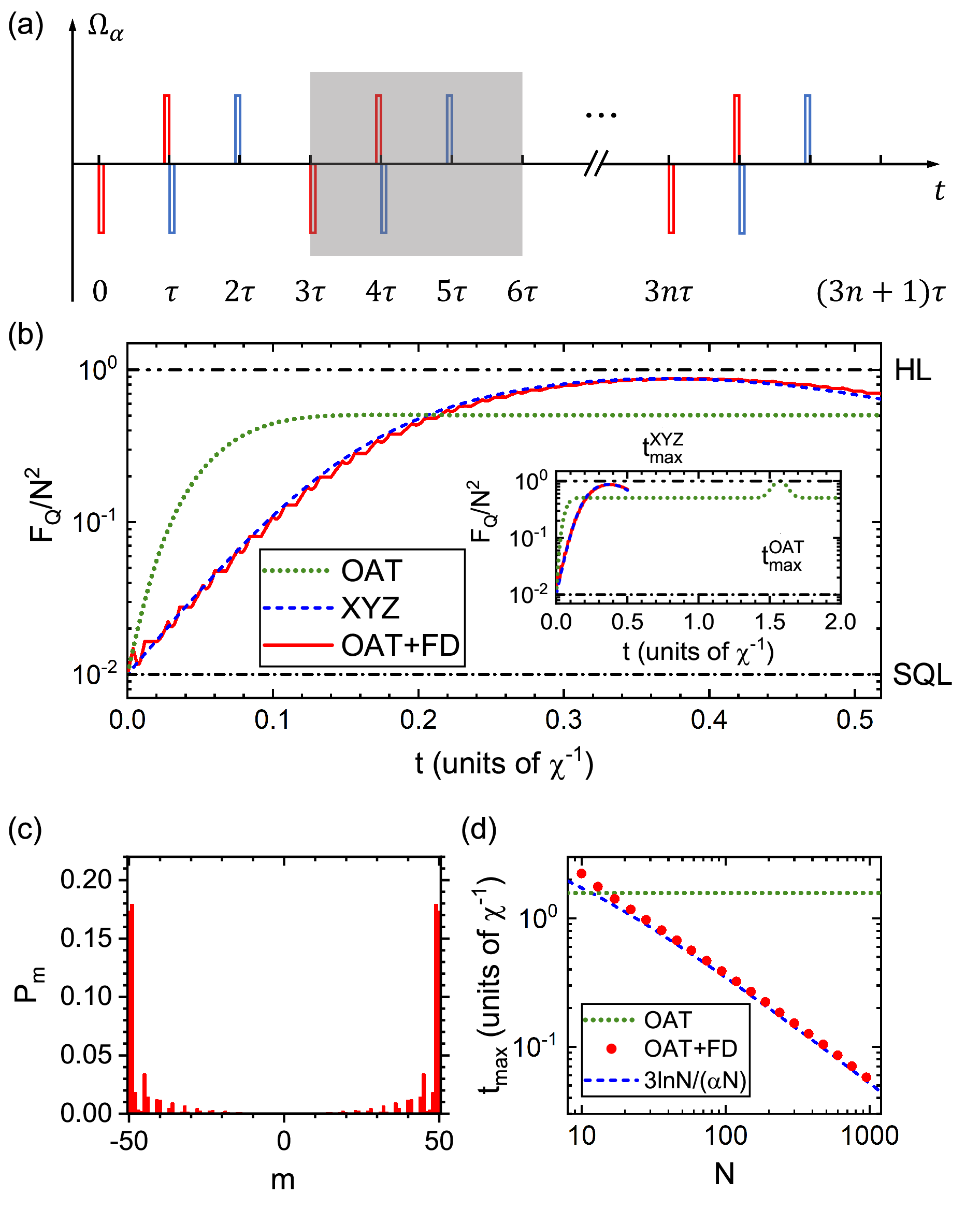}
    \caption{Generation of the catlike state with Floquet driving. (a) An
    illustration of the proposed pulse sequences. One period is $3\protect\tau $
    (shaded), consisting of $\pm \protect\pi /2$ pulses (up for
    \textquotedblleft $+$\textquotedblright\ and down for \textquotedblleft $-$%
    \textquotedblright ) along the $x$ axis (red one) and $y$ axis (blue one). (b)
    Time evolution of the optimal quantum Fisher information $F_{Q}$ of the
    original OAT model (green dotted), the effective XYZ model (blue dashed), and
    the proposed Floquet-driving scheme with driving parameter $\protect\alpha %
    =0.4$ (red solid) for $N=100$ particles. Horizontal lines indicate the
    standard quantum limit $F_{Q}=N$ (black dash-dotted line) and the Heisenberg
    limit $F_{Q}=N^{2}$ (black dash-dot-dotted line). The time when $F_{Q}$ reaches
    its maximum of the XYZ model and the OAT model are marked as $t_{\mathrm{max}%
    }^{\mathrm{XYZ}}$ and $t_{\mathrm{max}}^{\mathrm{OAT}}$, respectively. (c)
    Probability distribution $P_{m}=\abs{\braket{m}{\psi}}^{2}$ on eigenstates
    of $J_{y}$ at $t=t_{\mathrm{max}}^{\mathrm{XYZ}}$ for the XYZ model. (d)
    Optimal time $t_{\mathrm{max}}$ when $F_{Q}$ reaches its maximum for the OAT
    model (green dotted) and the proposed Floquet-driving scheme (red circles)
    versus particle number $N$, compared with the predicted value $t_{c}\simeq
    3\ln N/(\protect\alpha N)$ (blue dashed line). }
    \label{fig:2}
    \end{figure}
    
    The proposed XYZ model may be difficult to find directly in existing
    systems, as the collective-spin cubic Hamiltonian requires three-particle
    interaction. In light of this, we propose a Floquet-driving scheme to
    synthesize the XYZ model from an original OAT interaction $H_{\mathrm{OAT}%
    }=\chi J_{z}^{2}$ which commonly exists in many realistic systems \cite%
    {Gross2010,Riedel2010,Leroux2010,Matthew2018,Hines2023}. The scheme utilizes
    periodic $\pm \pi /2$ pulses along the $\alpha $ axis with $R_{\pm \pi
    /2}^{\alpha }=e^{\mp i\pi J_{\alpha }/2}$ ($\alpha =x,y$). Each period is $3\tau $ and is made up of the
    following: two $\pm \pi /2$
    rotations about $x$ axis separated by a free evolution for $\tau $, two $\pm \pi /2$ rotations about $y$
    axis separated by a second free evolution for $\tau $, and a third free evolution for $\tau $. This is a special case of the general time-dependent Hamiltonian $H(t)=\Omega_x(t)J_x+\Omega_y(t)J_y+H_\mathrm{OAT}$ with $\Omega_\alpha(t)$ consisting of a series of Dirac delta functions, which means the duration of driving pulses is short enough to be ignored, as illustrated in
    Fig.~\ref{fig:2}(a).
    The evolution operator for one period can be written as
    \begin{eqnarray}
    U &=&e^{-i\chi J_{z}^{2}\tau }R_{\pi /2}^{y}e^{-i\chi J_{z}^{2}\tau }R_{-\pi
    /2}^{y}R_{\pi /2}^{x}e^{-i\chi J_{z}^{2}\tau }R_{-\pi /2}^{x}  \nonumber \\
    &=&e^{-i\chi J_{z}^{2}\tau }e^{-i\chi J_{x}^{2}\tau }e^{-i\chi J_{y}^{2}\tau
    }.
    \end{eqnarray}%
    We can find that the driving pulses transform the original one-axis twisting
    into sequential three-axis twisting. Although the first order terms vanish
    as $J_{x}^{2}+J_{x}^{2}+J_{x}^{2}=j(j+1)$ is a constant, the second order
    terms are not canceled due to the noncommutativity of the operators. We
    simplify the expression using the Baker-Campbell-Hausdorff formula, which
    results in \cite{Supp}
    \begin{equation}
    U=\exp \left[ -i\chi ^{2}\tau ^{2}(J_{x}J_{y}J_{z}+J_{z}J_{y}J_{x})+\mathcal{%
    O}(\tau ^{2})\right] .
    \end{equation}%
    For small $\tau $ ($\chi \tau N/2\ll 1$), we can ignore the higher order
    terms and arrive at $U\simeq e^{-i\chi ^{2}\tau
    ^{2}(J_{x}J_{y}J_{z}+J_{z}J_{y}J_{x})}$, and the corresponding effective
    Hamiltonian becomes
    \begin{equation}
    H_{\mathrm{eff}}=\frac{1}{3}\chi ^{2}\tau (J_{x}J_{y}J_{z}+J_{z}J_{y}J_{x}),
    \end{equation}%
    which has exactly the form of the XYZ model \eqref{XYZ}, with effective
    interaction strength $\chi _{\mathrm{XYZ}}^{\mathrm{eff}}=\chi ^{2}\tau N/6$%
    . To safely drop higher order terms, we may set $\chi \tau =2\alpha /N$ with
    $\alpha \lesssim 1$ being an alternative driving parameter. The effective
    interaction strength then reads $\chi _{\mathrm{XYZ}}^{\mathrm{eff}}=\alpha
    \chi /3$. We mention that an interaction strength with opposite sign can be
    obtained by simply adjusting the order of pulse sequences.
    
    
    To verify the validity of the Floquet-driving scheme, in Fig.~\ref{fig:2}(b)
    we plot the time evolution of the optimal quantum Fisher information (QFI)
    in the Floquet-driving scheme, compared with the dynamics of the original
    OAT model and the effective XYZ model. 
    Assuming a phase $\phi $ is
    encoded via $e^{-iJ_{\mathbf{n}}\phi }$ to an input state $\hat{\rho}_{0}$,
    the QFI can be expressed as
    \begin{equation}
    F_{Q}(\hat{\rho}_{0},J_{\mathbf{n}})=2\sum_{q_{\kappa }+q_{\kappa ^{\prime
    }}>0}\frac{(q_{\kappa }-q_{\kappa ^{\prime }})^{2}}{q_{\kappa }+q_{\kappa
    ^{\prime }}}\abs{\bra{\kappa'}J_\mathbf{n}\ket{\kappa}}^{2},
    \end{equation}%
    where $\hat{\rho}_{0}=\sum_{\kappa }q_{\kappa }\ket{\kappa}\bra{\kappa}$
    represents the spectral decomposition of the input state. The 
    QFI is bounded by the Heisenberg limit $F_{Q}(\ket{\hat{\rho}_0},J_{\mathbf{n%
    }})\leq N^{2}$ for $N$ particles. For a pure input state $\hat{\rho}_{0}=%
    \ket{\psi_0}\bra{\psi_0}$, the expression is reduced to $F_{Q}(\ket{\psi_0}%
    ,J_{\mathbf{n}})=4(\Delta J_{\mathbf{n}})^{2}.$ 
    From Fig.~\ref{fig:2}(b) we can
    find that for the Floquet-driving scheme, the QFI can approach the
    Heisenberg limit in a fairly short time, in good accordance with the result
    of the effective XYZ model. Compared with the OAT dynamics which achieves
    the Heisenberg limit at $\chi t=\pi /2$, the time required in our scheme is
    greatly shortened.
    
    At $t=t_{\mathrm{max}}^{\mathrm{XYZ}}$ when $F_{Q}$ achieves its maximum for
    the XYZ model, the probability distribution on eigenstates of $J_{y}$ ($\{%
    \ket{m}_{y}\}_{m=-N/2}^{N/2}$, $J_{y}\ket{m}_{y}=m\ket{m}_{y}$) is shown in
    Fig.~\ref{fig:2}(c). We find that the distribution is perfectly symmetric
    for $\pm m$, with about $70\%$ of the population on $m=\pm N/2$ and $\pm
    (N/2-1)$, which resembles the GHZ state to a large extent. Note that this
    distribution is much more concentrated than the state created
    through twist-and-turn dynamics \cite{Supp}. The required generation time for different particle numbers is plotted in Fig.~\ref{fig:2}%
    (d). The result matches well with the predicted value $t_{c}\simeq \ln {N}%
    /(\chi _{\mathrm{eff}}N)=3\ln {N}/(\alpha N\chi )$. 
    
    \begin{figure}[tbp]
    \includegraphics[width=\columnwidth]{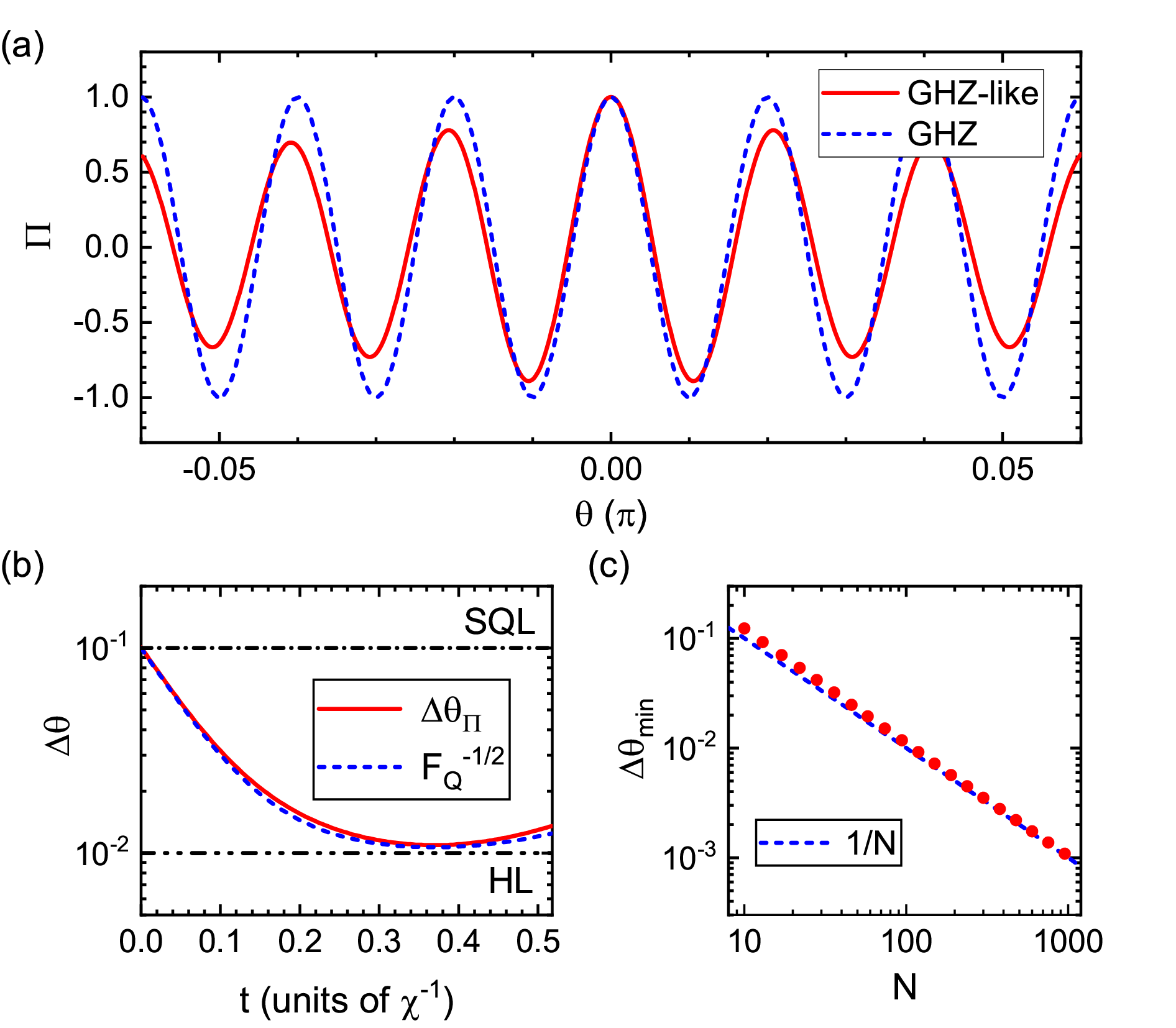}
    \caption{Parity oscillation of the GHZ-like state for $N=100$ particles. (a)
    The expectation value of parity $\Pi $ versus the azimuth angle $\protect%
    \theta $ of the rotation axis, for the GHZ-like state generated through the
    XYZ dynamics (red solid line) and the perfect GHZ state (black dashed line). (b) The
    precision $\Delta \protect\theta $ at $\protect\theta _{0}=\protect\pi /(2N)$
    versus evolution time $t$ of the XYZ dynamics (red solid line), in comparison
    with the quantum Cram\'{e}r-Rao bound $F_{Q}^{-1/2}$ (blue dashed line). (c) The
    optimal precision $\Delta \protect\theta _{\mathrm{min}}$ versus particle
    number $N$ of the GHZ-like state (red circles), comparing with the
    Heisenberg limit $1/N$ (blue dashed line).}
    \label{fig:3}
    \end{figure}
    
    A characteristic behavior of the GHZ state is the fast oscillation of
    parity \cite{Bollinger1996,Sackett2000,Leibfried2004,Omran2019}, defined as $%
    \Pi =\sum_{m}(-1)^{m}\ket{m}\bra{m}$ ($\ket{m}$ is the Dicke state with $J_{z}%
    \ket{m}=m\ket{m}$), after applying a $\pi /2$ rotation $e^{i\pi J_{\theta
    }/2}$ where $J_{\theta }=J_{y}\cos \theta -J_{x}\sin \theta $. This property
    is also possessed by the GHZ-like state generated through the XYZ dynamics.
    In Fig.~\ref{fig:3}(a) we plot the expectation value of parity with respect
    to $\theta $. 
    For a perfect GHZ state $%
    \ket{\mathrm{GHZ}}=(\ket{N/2}+\ket{-N/2})/\sqrt{2}$, the parity oscillation
    can be derived analytically as $\Pi (\theta )=\cos (N\theta )$. For the
    generated GHZ-like state [concentrating at $(0,0,\pm 1)$ on the Bloch sphere], we can see that the result resembles that of the
    GHZ state, verifying its GHZ-like property. 
    The components other than $m=\pm N/2$ in the GHZ-like state will degrade the amplitude of oscillation only when $\theta $ becomes larger.
    
    The high-frequency oscillation of parity can be applied in high-precision
    measurement to obtain a precision approaching the Heisenberg limit $\Delta
    \theta _{\mathrm{HL}}=1/N$. We choose $\theta _{0}=\pi /(2N)$ where the
    gradient $\abs{\partial_\theta\Pi}$ nearly reaches its maximum and calculate
    the precision as $\Delta \theta =\abs{\Delta\Pi/\partial_\theta\Pi}_{\theta
    =\theta _{0}}$. Figure~\ref{fig:3}(b) shows the result obtained with an
    input state which evolves under $H_{\mathrm{XYZ}}$ for $t$ from an initial
    coherent spin state $\ket{x}$. We find the precision is always
    approaching the quantum Cram\'{e}r-Rao bound $\Delta \theta _{\mathrm{QCRB}%
    }=F_{Q}^{-1/2}$, which means the parity measurement is always an ideal one
    to exploit the metrology-enhanced property generated by the XYZ model \cite%
    {Supp}. The optimal precision $\Delta \theta _{\mathrm{min}}$ versus
    particle number $N$ is plotted in Fig.~\ref{fig:3}(c), proving a
    Heisenberg-limited precision.
    
    \begin{figure}[tbp]
    \includegraphics[width=\columnwidth]{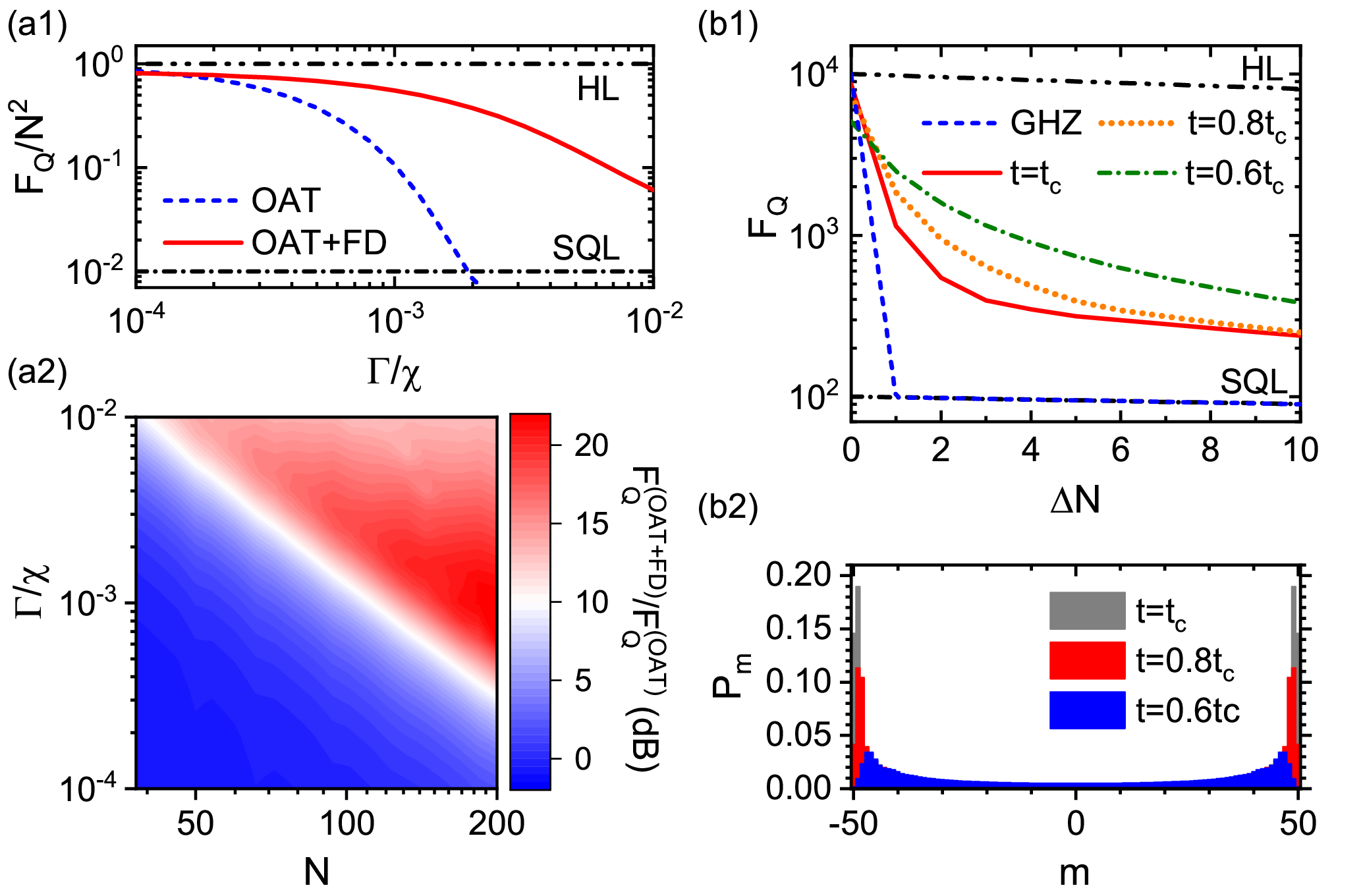}
    \caption{The effective XYZ dynamics and the obtained GHZ-like state in the
    presence of decoherence and losses. (a1) The quantum Fisher information $%
    F_{Q}$ of the GHZ(-like) state created through our Floquet-driving scheme
    (red solid) and OAT dynamics (blue dashed) for $N=100$ particles with
    respect to $\Gamma $. (a2) The ratio of the quantum Fisher information for
    two dynamics $F_{Q}^{\mathrm{(OAT+FD)}}/F_{Q}^{\mathrm{(OAT)}}$ versus $N$
    and $\Gamma $. Data is calculated at $\protect\chi t=3\ln N/(\protect\alpha %
    N)$ with $\protect\alpha =0.4$ for Floquet-driving scheme and $\protect\chi %
    t=\protect\pi /2$ for OAT scheme. (b1) The optimal quantum Fisher
    information of the perfect GHZ state (blue dashed) and the GHZ-like state
    generated through XYZ dynamics at $t=t_{c}$ (red solid), $t=0.8t_{c}$
    (orange dotted) and $t=0.6t_{c}$ (green dash-dotted) when $\Delta N$
    particles are lossed. (b2) The probability distribution $P_{m}$ for the
    GHZ-like state generated through XYZ dynamics at $t=t_{c}$ (grey), $%
    t=0.8t_{c}$ (red) and $t=0.6t_{c}$ (blue).}
    \label{fig:4}
    \end{figure}
    
    To analyze the impact of decoherence, we consider the effective OAT
    interaction mediated by an optical cavity as proposed in Ref. \cite%
    {Matthew2018}. In such schemes the cavity mode couples the ground $%
    \ket{\downarrow}\equiv {^{1}\mathrm{S}_{0}}$ and excited $\ket{\uparrow}%
    \equiv {^{3}\mathrm{P}_{0}}$ clock states of $^{87}\mathrm{Sr}$ atoms,
    leading to the effective OAT Hamiltonian $H_{\mathrm{eff}}=\chi
    (J^{2}-J_{z}^{2})$, accompanied with an effective Lindblad operator $L=%
    \sqrt{\Gamma /2}J_{-}$ which describes the superradiance caused by the
    leakage of photons out of the cavity. In Figs.~\ref{fig:4}(a1) and \ref{fig:4}(a2) we examine
    the effects of this typical decoherence channel on the creation of
    GHZ(-like) states. Since our Floquet-driving scheme greatly shortens the
    evolution time, it is reasonable to find that it outperforms the OAT scheme
    as the intensity of superradiance $\Gamma $ gets larger. The ratio of the
    QFI for two schemes $F_{Q}^{\mathrm{OAT+FD}}/F_{Q}^{\mathrm{OAT}}$ is
    plotted in Fig.~\ref{fig:4}(a2), which further confirms the superiority of
    our scheme in the case of a large particle number, owing to the $\ln N/N$
    dependence of the evolution time. One may notice that for large $N$, this
    ratio slightly drops after its first rise as $\Gamma $ grows. This is
    simplify the consequence that the OAT dynamics has been almost completely
    destroyed by the severe effect of decoherence, making $F_{Q}^{\mathrm{OAT}}$
    nearly unchanged as $\Gamma $ continues to increase.
    
    The GHZ state is also known to be extremely fragile to losses, with its
    coherence completely destroyed when only one particle is lost, while our
    GHZ-like state is more robust. In Fig.~\ref{fig:4}(b1) we compare the
    performance of the GHZ-like states with the perfect GHZ state in the
    presence of particle losses. The degradation of the QFI for the GHZ-like
    states is slighter than that of the perfect GHZ state, which can be further
    improved by shortening the evolution time and using a GHZ-like state with a
    wider probability distribution, as illustrated by the examples 
    of $t=0.8t_{c}$ and $0.6t_{c}$ in Figs.~\ref{fig:4}(b1) and \ref{fig:4}(b2). This can be 
    interpreted as the wider distribution ensuring more 
    coherence lying in the matrix elements $\ket{\pm(N/2-n)}\bra{\mp(N/2-n)}$, 
    which survives after $n\geq1$ particles are lost. We note that the QFI of the
    GHZ-like states seems to asymptotically stable to a fixed value above the
    standard quantum limit when the number of lost particles $\Delta N$ is
    large, indicating that each one of the two parts of the GHZ-like state is
    still nonclassical and possesses metrologically useful properties.
    
    In conclusion, we present a three-body collective-spin XYZ model to rapidly
    create GHZ-like states. It shows a significant shortcut of timescale from $%
    \chi t=\pi /2$ of the OAT model to $\chi t\sim \ln {N}/N$ of our model. We
    proposed to realize the XYZ model by employing Floquet driving to an
    original OAT interaction. We
    further study the behavior of GHZ-like states in parity measurements, and
    obtain a Heisenberg-limited level of measurement precision, showing that the
    metrology-enhanced property of the XYZ model can be always fully exploited
    by taking parity measurements. Moreover, we investigate the behavior of our
    scheme and the obtained GHZ-like state in the presence of decoherence and
    losses, finding it more robust than the OAT scheme and the ideal GHZ state.
    Our work paves the way for the creation of large-particle-number GHZ-like
    states, which has significant applications in quantum metrology and quantum
    information science.
    
    \begin{acknowledgments}
    This work is supported by the National Key R\&D Program of China (Grant No. 2023YFA1407600), and the National Natural Science Foundation of China (NSFC) (Grants No. 12275145, No. 92050110, No. 91736106, No. 11674390, and No. 91836302).
    \end{acknowledgments}
    \bibliography{catlike_main}
\begin{widetext}

\begin{center}
\textbf{\Large Supplemental Material}
\end{center}

\section{Semi-classical treatment of the XYZ model}

In this section we apply the semi-classical treatment to explore the
metrological properties of the XYZ model
\begin{equation}\label{Hxyz}
H_{\mathrm{XYZ}}=\frac{2\chi _{\mathrm{xyz}}}{N}%
(J_{x}J_{y}J_{z}+J_{z}J_{y}J_{x}).
\end{equation}%
Utilizing the commutations $\comm{J_\alpha}{J_\beta}=i\epsilon _{\alpha
\beta \gamma }J_{\gamma }$ ($\alpha ,\beta ,\gamma =x,y,z$), it can also be
written as
\begin{equation}
H_{\mathrm{XYZ}}=\frac{2\chi _{\mathrm{xyz}}}{N}%
(J_{y}J_{x}J_{z}+J_{z}J_{x}J_{y})=\frac{2\chi _{\mathrm{xyz}}}{N}%
(J_{x}J_{z}J_{y}+J_{y}J_{z}J_{x}).
\end{equation}

We write the Heisenberg equation of motion for the expectation values of spin
components, that is $d\expval{J_\alpha}/dt=-i\expval{\comm{J_\alpha}{H_%
\mathrm{XYZ}}}$. Defining varibles of the semi-classical phase space $%
(X,Y,Z)=(\expval{J_x,J_y,J_z})/j$, where $j=N/2$ and $X^2+Y^2+Z^2=1$, the
classical flow in the phase space is described by
\begin{equation}
\frac{d}{dt} \begin{pmatrix} X \\ Y \\ Z \\ \end{pmatrix} =\chi_\mathrm{xyz}%
N \begin{pmatrix} (Z^2-Y^2)X \\ (X^2-Z^2)Y \\ (Y^2-X^2)Z \\ \end{pmatrix},
\end{equation}
which is obtained by neglecting correlations (setting $\expval{J_\alpha
J_\beta}=\expval{J_\alpha}\expval{J_\beta}$).

There are 6 saddle points in the phase space: $(\pm 1,0,0)$, $(0,\pm 1,0)$
and $(0,0,\pm 1)$, being able to generate metrologically useful states.
Since they are all equivalent due to the symmetry of the Hamiltonian under
cyclic permutations, we focus on one point $(1,0,0)$ and examine the motion
of a initial coherent spin state centered at it ($\ket{x}$ with $J_{x}\ket{x}%
=j\ket{x}$). For $\chi _{\mathrm{xyz}}>0$, the variance is found to be
suppressed along $z$ axis and stretched along $y$ axis. The points on the
great circle defined by $z=0$ will move along it, so that we have $X=\sqrt{%
1-Y^{2}}$ and the equation of motion becomes
\begin{equation}
\frac{dY}{dt}=\chi _{\mathrm{xyz}}N(1-Y^{2})Y,  \label{dYdt}
\end{equation}%
resulting in
\begin{equation}
\chi _{\mathrm{xyz}}t_{c}=\frac{1}{N}\int_{Y_{0}}^{Y_{f}}\frac{dY}{Y(1-Y^{2})%
}.
\end{equation}

To find the time to obtain the GHZ-like state, we use the method proposed in
Ref.~\cite{Munoz2023}, where we expect the point initially at the edge of
the initial uncertainty patch $(\sqrt{1-1/N},\pm 1/\sqrt{N},0)$ will end at $%
(1/\sqrt{N},\pm \sqrt{1-1/N},0)$, the uncertainty edge of a coherent spin
state centered at $(0,\pm 1,0)$, as illustrated in Fig.~\ref{fig:S1}. As a
result, we set $Y_{0}=1/\sqrt{N}$ and $Y_{f}=\sqrt{1-1/N}$. The time to
generate the GHZ-like state is then estimated as
\begin{equation}
\chi _{\mathrm{xyz}}t_{c}=\frac{1}{N}\ln (N-1)\simeq \frac{\ln N}{N},
\label{tc_classical}
\end{equation}%
which is valid for $N\gg 1$.

\begin{figure}[tbp]
\includegraphics[width=0.9\columnwidth]{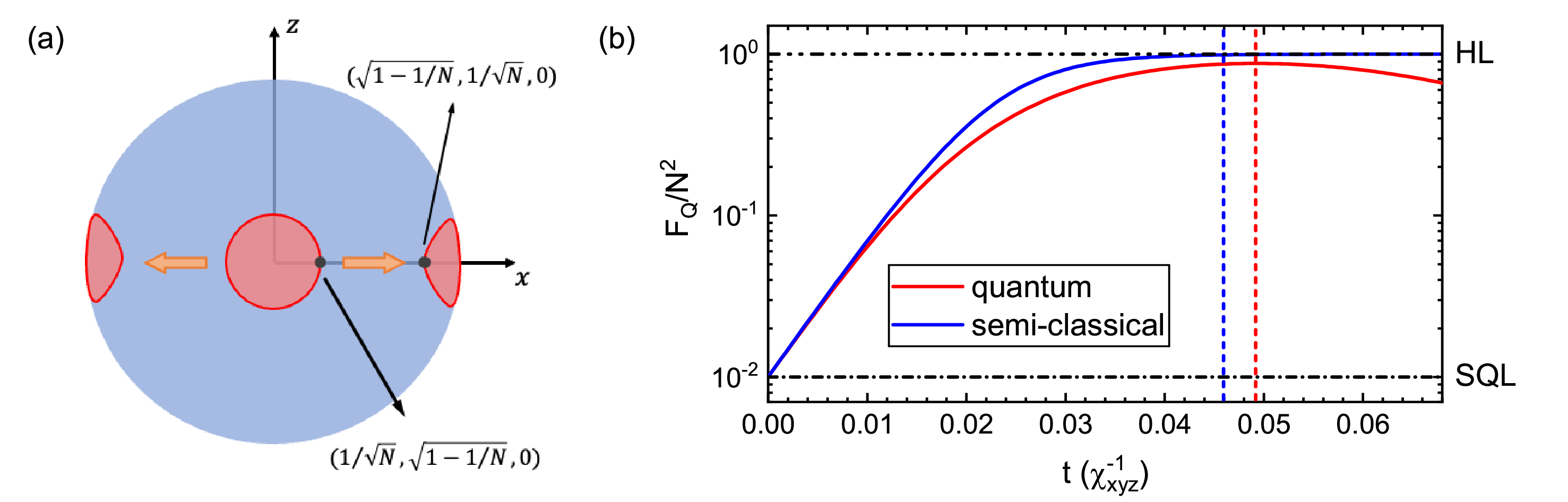}
\caption{(a) illustration of the evolution from a initial coherent spin
state to a final GHZ-like state in the semi-classical picture. (b) Time
evolution of the quantum Fisher information $F_Q$ under XYZ dynamics for $%
N=100$ particles. We compare the exact numerical result (red solid) with the
semi-classical result \eqref{Fq_classical} (blue solid). The exact optimal
time and that obtained from semi-classical treatment \eqref{tc_classical}
are labeled with red and blue dashed lines, respectively.}
\label{fig:S1}
\end{figure}

We can further derive the explicit expression for the time evolution of the
quantum Fisher information within the semi-classical description. Since the
variance is stretched along $y$ axis, we take $J_y$ as the generator and get
\begin{equation}
F_Q = 4\left(\Delta J_y\right)^2 = N^2 Y^2.
\end{equation}
From \eqref{dYdt} we obtain
\begin{equation}
\frac{dY^2}{dt}=2Y\frac{dY}{dt}=2\chi_\mathrm{xyz}N(1-Y^2)Y^2,
\end{equation}
which leads to
\begin{equation}
Y^2(t) = \frac{1}{1+\frac{1-Y_0^2}{Y_0^2}e^{-2\chi_\mathrm{xyz}Nt}}.
\end{equation}
Recalling that $Y_0=1/\sqrt{N}$, we get
\begin{equation}  \label{Fq_classical}
F_Q=N^2Y^2=\frac{N^2}{1+(N-1)e^{-2\chi_\mathrm{xyz}Nt}}.
\end{equation}

We may examine the validity of the expression \eqref{Fq_classical} by
comparing it with the exact numerical result, as shown in Figure~\ref{fig:S1}%
. The semi-classical result matches well with the exact one before $t=t_c$.
The optimal time given by \eqref{tc_classical} is also closed to the exact
result.

\section{The Floquet-driving scheme to realize the XYZ model}

As shown in Fig.~2(a) of the main text, we apply periodic $\pm\pi/2$ pulses to the original
one-axis twisting (OAT) interaction
\begin{equation}
H_\mathrm{OAT}=\chi J_z^2.
\end{equation}
The evolution operator for one period is given by
\begin{eqnarray}
U &=& e^{-i\chi J_z^2\tau}e^{-i\frac{\pi}{2}J_y}e^{-i\chi J_z^2\tau}e^{i%
\frac{\pi}{2}J_y}e^{-i\frac{\pi}{2}J_x}e^{-i\chi J_z^2\tau}e^{i\frac{\pi}{2}%
J_x}  \nonumber \\
&=& e^{-i\chi J_z^2\tau}e^{-i\chi J_x^2\tau}e^{-i\chi J_y^2\tau},
\end{eqnarray}
where we have used $e^{-i\pi J_{y(x)}/2}J_ze^{i\pi J_{y(x)}/2}=(-)J_{x(y)}$.

To derive the effective Hamiltonian, we introduce the
Baker-Campbell-Hausdorff (BCH) formula
\begin{equation}
e^{A}e^{B}=\exp (A+B+\frac{1}{2}\comm{A}{B}+\frac{1}{12}\comm{A}{\comm{A}{B}}%
-\frac{1}{12}\comm{B}{\comm{A}{B}}+...).
\end{equation}%
The product of the first two terms is calculated by setting
\begin{equation}
A=-i\chi J_{z}^{2}\tau ,\quad B=-i\chi J_{x}^{2}\tau .
\end{equation}%
Assuming $\tau $ is small, the first and second orders of the BCH formula
are
\begin{eqnarray}
A+B &=&-i\chi (J_{z}^{2}+J_{x}^{2})\tau =-i\chi (\mathbf{J}%
^{2}-J_{y}^{2})\tau ,  \nonumber \\
\frac{1}{2}\comm{A}{B} &=&-\frac{1}{2}\chi ^{2}\tau ^{2}\comm{J_z^2}{J_x^2}%
=-i\chi ^{2}\tau ^{2}(J_{x}J_{y}J_{z}+J_{z}J_{y}J_{x}),
\end{eqnarray}%
and the third order is
\begin{equation}
\frac{1}{12}\comm{A}{\comm{A}{B}}-\frac{1}{12}\comm{B}{\comm{A}{B}}=\frac{i}{%
12}\chi ^{3}\tau ^{3}\left( \comm{J_z^2}{\comm{J_z^2}{J_x^2}}-%
\comm{J_x^2}{\comm{J_z^2}{J_x^2}}\right) .
\end{equation}%
So the evolution operator is reduced to
\begin{equation}
U=\exp \left[ -i\chi \tau (\mathbf{J}^{2}-J_{y}^{2})-i\chi ^{2}\tau
^{2}(J_{x}J_{y}J_{z}+J_{z}J_{y}J_{x})+\frac{i}{12}\chi ^{3}\tau ^{3}\left( %
\comm{J_z^2}{\comm{J_z^2}{J_x^2}}-\comm{J_x^2}{\comm{J_z^2}{J_x^2}}\right)
+...\right] e^{-i\chi J_{y}^{2}\tau }.
\end{equation}%
Applying the BCH formula again, the first order term is
\begin{equation}
-i\chi \tau (\mathbf{J}^{2}-J_{y}^{2})-i\chi \tau J_{y}^{2}=-i\chi \tau
\mathbf{J}^{2},
\end{equation}%
which is a constant and can be neglected. The first nonvanishing term is
\begin{equation}
-i\chi ^{2}\tau ^{2}(J_{x}J_{y}J_{z}+J_{z}J_{y}J_{x}),
\end{equation}%
and the third order is
\begin{eqnarray}
&&\frac{i}{12}\chi ^{3}\tau ^{3}\left( 6\comm{\comm{J_z^2}{J_x^2}}{J_y^2}+%
\comm{J_z^2}{\comm{J_z^2}{J_x^2}}-\comm{J_x^2}{\comm{J_z^2}{J_x^2}}\right)
\nonumber \\
&=&\frac{i}{12}\chi ^{3}\tau ^{3}\left( 7\comm{J_z^2}{\comm{J_z^2}{J_x^2}}+5%
\comm{J_x^2}{\comm{J_z^2}{J_x^2}}\right)   \nonumber \\
&=&-\frac{i}{6}\chi ^{3}\tau ^{3}\left[ 7\left(
J_{z}^{2}J_{y}^{2}+J_{y}^{2}J_{z}^{2}\right) -2\left(
J_{z}^{2}J_{x}^{2}+J_{x}^{2}J_{z}^{2}\right) -5\left(
J_{x}^{2}J_{y}^{2}+J_{y}^{2}J_{x}^{2}\right) +14J_{z}\left(
J_{y}^{2}-J_{x}^{2}\right) J_{z}+10J_{x}\left( J_{z}^{2}-J_{y}^{2}\right)
J_{x}\right] .
\end{eqnarray}%
To safely drop the third and higher order terms, we require
\begin{equation}
\chi \tau \frac{N}{2}\ll 1.  \label{req_chi}
\end{equation}%
The evolution operator now takes the form
\begin{equation}
U=\exp \left[ -i\chi ^{2}\tau ^{2}\left(
J_{x}J_{y}J_{z}+J_{z}J_{y}J_{x}\right) +\mathcal{O}(\tau ^{2})\right] .
\end{equation}%
We can rewrite it in the form of the effective Hamiltonian
\begin{equation}
U(3\tau )=\exp \left( -i3\tau H_{\mathrm{eff}}\right) ,
\end{equation}%
leading to
\begin{equation}
H_{\mathrm{eff}}=\frac{1}{3}\chi ^{2}\tau \left(
J_{x}J_{y}J_{z}+J_{z}J_{y}J_{x}\right) ,
\end{equation}%
which is the XYZ Hamiltonian with the effective interaction strength
\begin{equation}
\chi _{\mathrm{xyz}}^{\mathrm{eff}}=\frac{1}{6}\chi ^{2}\tau N.
\end{equation}

The effective interaction strength is $\propto\tau$, which means we can
shorten the time to produce the GHZ-like state by using a large $\tau$.
However, the requirement \eqref{req_chi} may be violated as $\tau$ gets
larger, leading to the destruction of our scheme. We define an alternative
driving parameter
\begin{equation}
\alpha = \chi\tau\frac{N}{2},
\end{equation}
which characterizes the validity of the scheme (a smaller $\alpha$
corresponds to a more valid scheme).

\begin{figure}[tbp]
\includegraphics[width=0.9\columnwidth]{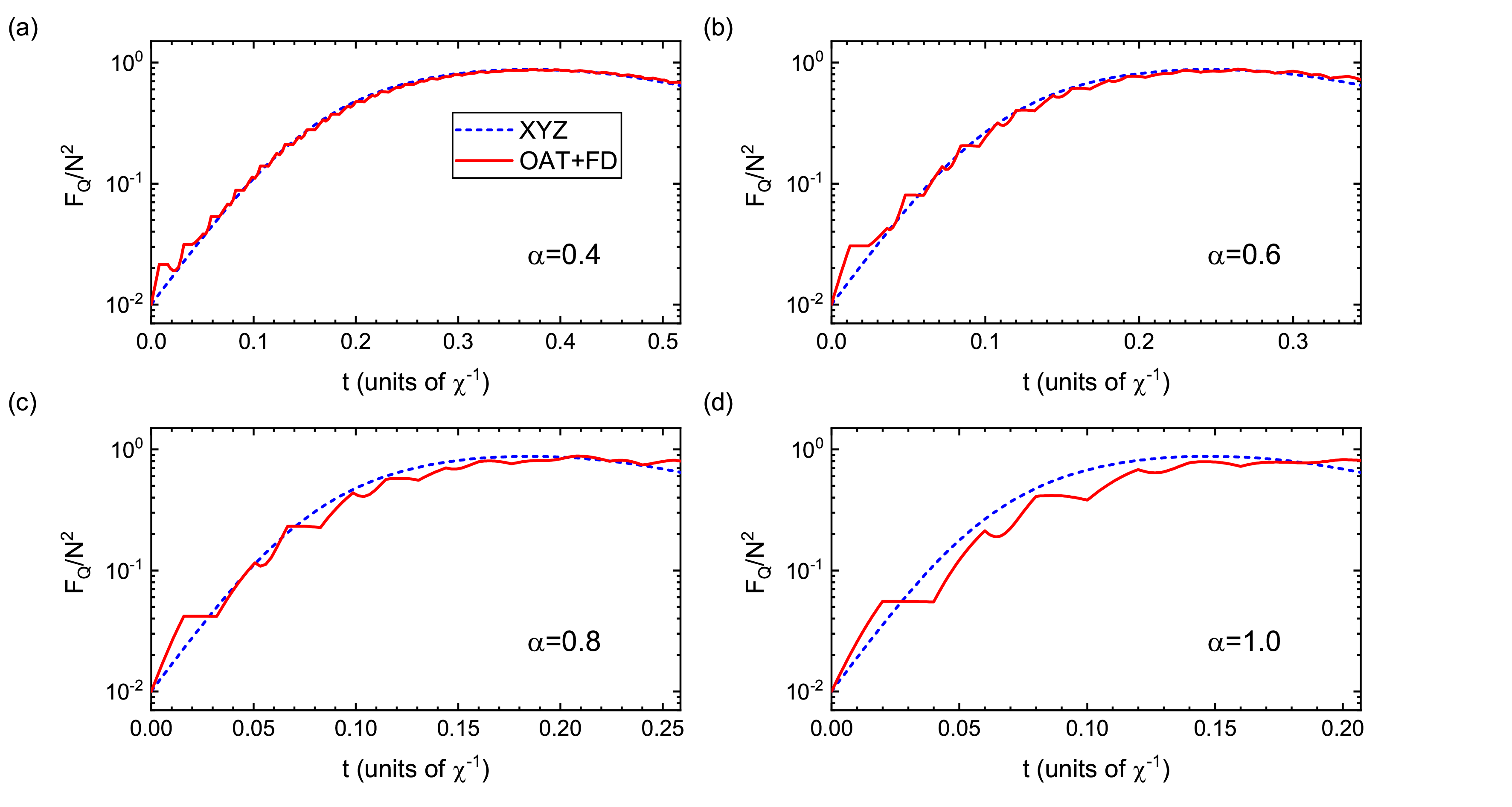}
\caption{Time evolution of the quantum Fisher information $F_Q$ for the
effective XYZ model (blue dashed) and the Floquet-driving scheme (red
solid), with driving parameter $\protect\alpha=$ (a) $0.4$, (b) $0.6$, (c) $%
0.8$, (d) $1.0$. The particle number $N=100$.}
\label{fig:S2}
\end{figure}

From Figure~\ref{fig:S2} we find that although the result matches better
with smaller $\alpha$, the optimal point can still be nearly achieved even
with $\alpha=1$, which obviously violates \eqref{req_chi}. This will largely
reduce the number of needed pulses and shorten the time to generate the
GHZ-like state, thus is favorable for experimental realization.

\section{Parity oscillation and metrological property of the GHZ-like states}

The high-frequency oscillation of the parity is often used to verify a GHZ
state. In this section we will deduce the expression of the expectation
value of parity for the GHZ-like state, showing its analogy and distinction
with the perfect GHZ state.

Before the derivation, first we prove the GHZ-like state can be expressed as
(consider $N$ is even)
\begin{equation}  \label{GHZ-like}
\ket{\mathrm{GHZ\text{-}like}} = d_0\ket{0} + \sum_{m>0}^{N/2}d_m\left(%
\ket{m}+\ket{-m}\right),
\end{equation}
where $\ket{m}$ is the eigenstate of $J_z$ with $J_z\ket{m}=m\ket{m}$ and $%
d_m$ is a real number for all $m$. The normalization requires $%
d_0^2+2\sum_{m>0}^{N/2}d_m^2=1$. This property originates from the symmetry
of the XYZ Hamiltonian. We can see it by expanding the state vector as $%
\ket{\psi(t)}=\sum_{m=-N/2}^{N/2}c_m(t)\ket{m}$, and writing out the
explicit Schr\"{o}dinger equation
\begin{equation}
i\sum_m\dot{c}_m(t)\ket{m}=\frac{2\chi_\mathrm{xyz}}{N}%
\sum_m(J_xJ_yJ_z+J_zJ_yJ_x)c_m(t)\ket{m},
\end{equation}
where we consider the negative interaction strength case ($\chi_\mathrm{xyz}%
<0$). Comparing the coefficient of $\ket{m}$, we obtain
\begin{eqnarray}  \label{se}
\dot{c}_m(t) &=& \frac{\chi_\mathrm{xyz}}{N}(m-1)\sqrt{%
(j-m+1)(j-m+2)(j+m-1)(j+m)}c_{m-2}  \nonumber \\
&&-\frac{\chi_\mathrm{xyz}}{N}(m+1)\sqrt{(j-m-1)(j-m)(j+m+1)(j+m+2)}c_{m+2}.
\end{eqnarray}
Replacing $m$ with $-m$, we obtain
\begin{eqnarray}  \label{se2}
\dot{c}_{-m}(t) &=& \frac{\chi_\mathrm{xyz}}{N}(m-1)\sqrt{%
(j-m+1)(j-m+2)(j+m-1)(j+m)}c_{-(m-2)}  \nonumber \\
&&-\frac{\chi_\mathrm{xyz}}{N}(m+1)\sqrt{(j-m-1)(j-m)(j+m+1)(j+m+2)}%
c_{-(m+2)}.
\end{eqnarray}
The only difference between \eqref{se} and \eqref{se2} is $c_k
\leftrightarrow c_{-k}$. For a initial coherent spin state polarized along $x
$ axis $\ket{\psi(0)}=\ket{x}$, the expansion coefficients are given by
\begin{equation}
c_m(0) = \frac{1}{2^{N/2}}\sqrt{\frac{N!}{(\frac{N}{2}+m)!(\frac{N}{2}-m)!}},
\label{initial}
\end{equation}
which is completely symmetric for $\pm m$, thus we have $c_m(t) = c_{-m}(t)$
for all $m$ and $t$. Additionally, all parameters appear in \eqref{se} and %
\eqref{initial} are real, so that $c_m(t)$ must be real for all $m$ and $t$.
As a result, we obtain the expression \eqref{GHZ-like} by defining $%
d_m=c_m(t)$.

Now we move on to derive the expression of parity oscillation for the
GHZ-like state \eqref{GHZ-like}. The parity $\Pi=\sum_m (-1)^{m}\ket{m}%
\bra{m}$ is measured after applying a $\pi/2$ rotation $e^{-i\pi J_\theta/2}$
to the GHZ-like state, where $J_\theta=J_y\cos\theta-J_x\sin\theta$. We
define single-particle states $\ket{\uparrow}$ and $\ket{\downarrow}$ as
eigenstates of $\sigma_z$ with eigenvalues $+1$ and $-1$. The eigenstate of $%
J_z$ can be written as
\begin{equation}
\ket{m} = \frac{1}{\mathcal{N}(m)}\sum_{\mathcal{P}} \left[\ket{\uparrow}%
^{N/2+m}\ket{\downarrow}^{N/2-m}\right],
\end{equation}
where $\sum_{\mathcal{P}}$ represents the summation over all permutations of
$N$ particles and $\mathcal{N}(m)$ is the normalized coefficient. The parity
operator may be expressed as $\Pi=\prod_{k=1}^N \sigma_z^k$. The expectation
value we need to calculate is
\begin{eqnarray}  \label{Pi}
\expval{\Pi(\theta)} &=& \bra{\mathrm{GHZ\text{-}like}}e^{i\pi
J_\theta/2}\Pi e^{-i\pi J_\theta/2}\ket{\mathrm{GHZ\text{-}like}}  \nonumber
\\
&=& \frac{d_0^2}{\mathcal{N}(0)^2}\sum_{\mathcal{P},\mathcal{P}^{\prime
}}\left(\bra{\uparrow}^{N/2}\bra{\downarrow}^{N/2}\right)e^{i\pi
J_\theta/2}\Pi e^{-i\pi J_\theta/2}\left(\ket{\downarrow}^{N/2}\ket{\uparrow}%
^{N/2}\right)  \nonumber \\
&&+\sum_{m>0}\frac{d_0d_m}{\mathcal{N}(0)\mathcal{N}(m)}\sum_{\mathcal{P},%
\mathcal{P}^{\prime }}  \nonumber \\
&&\times\left\{\left(\bra{\uparrow _y}^{N/2}\bra{\downarrow}%
^{N/2}\right)e^{i\pi J_\theta/2}\Pi e^{-i\pi J_\theta/2}\left[\ket{\uparrow}%
^{N/2+m}\ket{\downarrow}^{N/2-m}+\ket{\uparrow}^{N/2-m}\ket{\downarrow}%
^{N/2+m}\right]+\mathrm{H}.\mathrm{C}.\right\}  \nonumber \\
&&+ \sum_{m,m^{\prime }>0}\frac{d_m d_{m^{\prime }}}{\mathcal{N}(m)\mathcal{N%
}(m^{\prime })}\sum_{\mathcal{P},\mathcal{P}^{\prime }}  \nonumber \\
&&\times \left[\bra{\uparrow}^{N/2+m^{\prime }}\bra{\downarrow}%
^{N/2-m^{\prime }}+\bra{\uparrow}^{N/2-m^{\prime }}\bra{\downarrow}%
^{N/2+m^{\prime }}\right]e^{i\pi J_\theta/2}  \nonumber \\
&&\times \prod_{k=1}^N\sigma_z^k e^{-i\pi J_\theta/2}\left[\ket{\uparrow}%
^{N/2+m}\ket{\downarrow}^{N/2-m}+\ket{\uparrow}^{N/2-m}\ket{\downarrow}%
^{N/2+m}\right].
\end{eqnarray}
We notice that
\begin{eqnarray}
e^{-i\pi(\sigma_y\cos\theta-\sigma_x\sin\theta)/4}\ket{\uparrow}&=&\frac{1}{%
\sqrt{2}}\left[1+i(\sigma_y\cos\theta-\sigma_x\sin\theta)\right]%
\ket{\uparrow}=\frac{1}{\sqrt{2}}\left(\ket{\uparrow}-e^{i\theta}%
\ket{\downarrow}\right),  \nonumber \\
e^{-i\pi(\sigma_y\cos\theta-\sigma_x\sin\theta)/4}\ket{\downarrow}&=&\frac{1%
}{\sqrt{2}}\left[1+i(\sigma_y\cos\theta-\sigma_x\sin\theta)\right]%
\ket{\downarrow}=\frac{1}{\sqrt{2}}\left(e^{-i\theta}\ket{\uparrow}+%
\ket{\downarrow}\right),
\end{eqnarray}
which can be used to simplify the expression \eqref{Pi} as
\begin{eqnarray}
&&\left(\bra{\uparrow _y}^{N/2}\bra{\downarrow}^{N/2}\right)e^{i\pi
J_\theta/2}\Pi e^{-i\pi J_\theta/2}\left(\ket{\downarrow _y}^{N/2}%
\ket{\uparrow}^{N/2}\right)  \nonumber \\
&=&\frac{1}{2^N}\left[\left(\bra{\uparrow}-e^{-i\theta}\bra{\downarrow}%
\right)^{N/2}\left(e^{i\theta}\bra{\uparrow}+\bra{\downarrow}\right)^{N/2}%
\right]\prod_{k=1}^N\sigma_z^k\left[\left(e^{-i\theta}\ket{\uparrow}+%
\ket{\downarrow _y}\right)^{N/2}\left(\ket{\uparrow}-e^{i\theta}%
\ket{\downarrow}\right)^{N/2}\right]  \nonumber \\
&=&\frac{1}{2^N}\left[\left(\bra{\uparrow}-e^{-i\theta}\bra{\downarrow}%
\right)^{N/2}\left(e^{i\theta}\bra{\uparrow}+\bra{\downarrow}\right)^{N/2}%
\right]\left[\left(e^{-i\theta}\ket{\uparrow}-\ket{\downarrow _y}%
\right)^{N/2}\left(\ket{\uparrow}+e^{i\theta}\ket{\downarrow}\right)^{N/2}%
\right]=1.
\end{eqnarray}
and
\begin{eqnarray}
&&\left[\bra{\uparrow}^{N/2+m^{\prime }}\bra{\downarrow}^{N/2-m^{\prime }}+%
\bra{\uparrow}^{N/2-m^{\prime }}\bra{\downarrow}^{N/2+m^{\prime }}\right]%
e^{i\pi J_\theta/2} \prod_{k=1}^N\sigma_z^k e^{-i\pi J_\theta/2}\left[%
\ket{\uparrow}^{N/2+m}\ket{\downarrow}^{N/2-m}+\ket{\uparrow}^{N/2-m}%
\ket{\downarrow}^{N/2+m}\right]  \nonumber \\
&=& \frac{1}{2^N}\left[\left(\bra{\uparrow}-e^{-i\theta}\bra{\downarrow}%
\right)^{\otimes(N/2+m^{\prime })}\left(e^{i\theta}\bra{\uparrow}+%
\bra{\downarrow}\right)^{\otimes(N/2-m^{\prime })}+\left(\bra{\uparrow}%
-e^{-i\theta}\bra{\downarrow}\right)^{\otimes(N/2-m^{\prime
})}\left(e^{i\theta}\bra{\uparrow}+\bra{\downarrow}\right)^{\otimes(N/2+m^{%
\prime })}\right]  \nonumber \\
&&\times\left[\left(\ket{\uparrow}+e^{i\theta}\ket{\downarrow}%
\right)^{\otimes(N/2+m)}\left(e^{-i\theta}\ket{\uparrow}-\ket{\downarrow}%
\right)^{\otimes(N/2-m)}+\left(\ket{\uparrow}+e^{i\theta}\ket{\downarrow}%
\right)^{\otimes(N/2-m)}\left(e^{-i\theta}\ket{\uparrow}-\ket{\downarrow}%
\right)^{\otimes(N/2+m)}\right]  \nonumber \\
&=& \frac{1}{2^N}\left[(2e^{-i\theta})^{N/2+m}(2e^{i\theta})^{N/2-m}+(2e^{-i%
\theta})^{N/2-m}(2e^{i\theta})^{N/2+m}\right]\delta_{m,m^{\prime
}}=2\cos(m\theta)\delta_{m,m^{\prime }}.
\end{eqnarray}
The result is identical for all simultaneous permutations of $N$ particles,
thus we get
\begin{equation}  \label{parity}
\expval{\Pi(\theta)}=d_0^2+\sum_{m>0}^{N/2} 2d_m^2\cos(2m\theta).
\end{equation}
We can always find
\begin{equation}
\expval{\Pi(0)}=d_0^2+\sum_{m>0}^{N/2}2d_m^2=1,
\end{equation}
due to the normalization of the GHZ-like state \eqref{GHZ-like}.

For the perfect GHZ state $\ket{\mathrm{GHZ}}=(\ket{N/2}+\ket{-N/2})/\sqrt{2}
$, we have $d_m^\mathrm{GHZ}=\delta_{m,N/2}/\sqrt{2}$, the expression is
\begin{equation}
\expval{\Pi(\theta)}_\mathrm{GHZ}=\cos(N\theta),
\end{equation}
which oscillates at frequency $N$, leading to a Heisenberg-limited noise
reduction $\Delta\theta\sim1/N$.

We then consider an approximate GHZ state $\ket{\mathrm{AGHZ}}=(\ket{N/2}+%
\ket{N/2-1}+\ket{-N/2+1}+\ket{-N/2})/2$, which is expressed by %
\eqref{GHZ-like} with $d_m^\mathrm{AGHZ}=(\delta_{m,N/2}+\delta_{m,N/2-1})/2$%
, resulting in
\begin{equation}
\expval{\Pi(\theta)}_\mathrm{AGHZ}=\frac{1}{2}\left\{\cos(N\theta)+\cos[%
(N-2)\theta]\right\}=\cos[(N-1)\theta]\cos\theta.
\end{equation}
This situation is closer to the optimal GHZ-like state generated through the
XYZ dynamics. The factor $\cos\theta$ appears due to the interference of two
different components $m=\pm N/2$ and $m=\pm (N/2-1)$. When focusing on small
$\theta$, this factor $\cos\theta\simeq1$, the parity oscillates
approximately $\propto\cos[(N-1)\theta]$. Thus it provides a precision as
high as $\Delta\theta\sim1/(N-1)$, approaching the Heisenberg limit.

\begin{figure}[tbp]
\includegraphics[width=0.9\columnwidth]{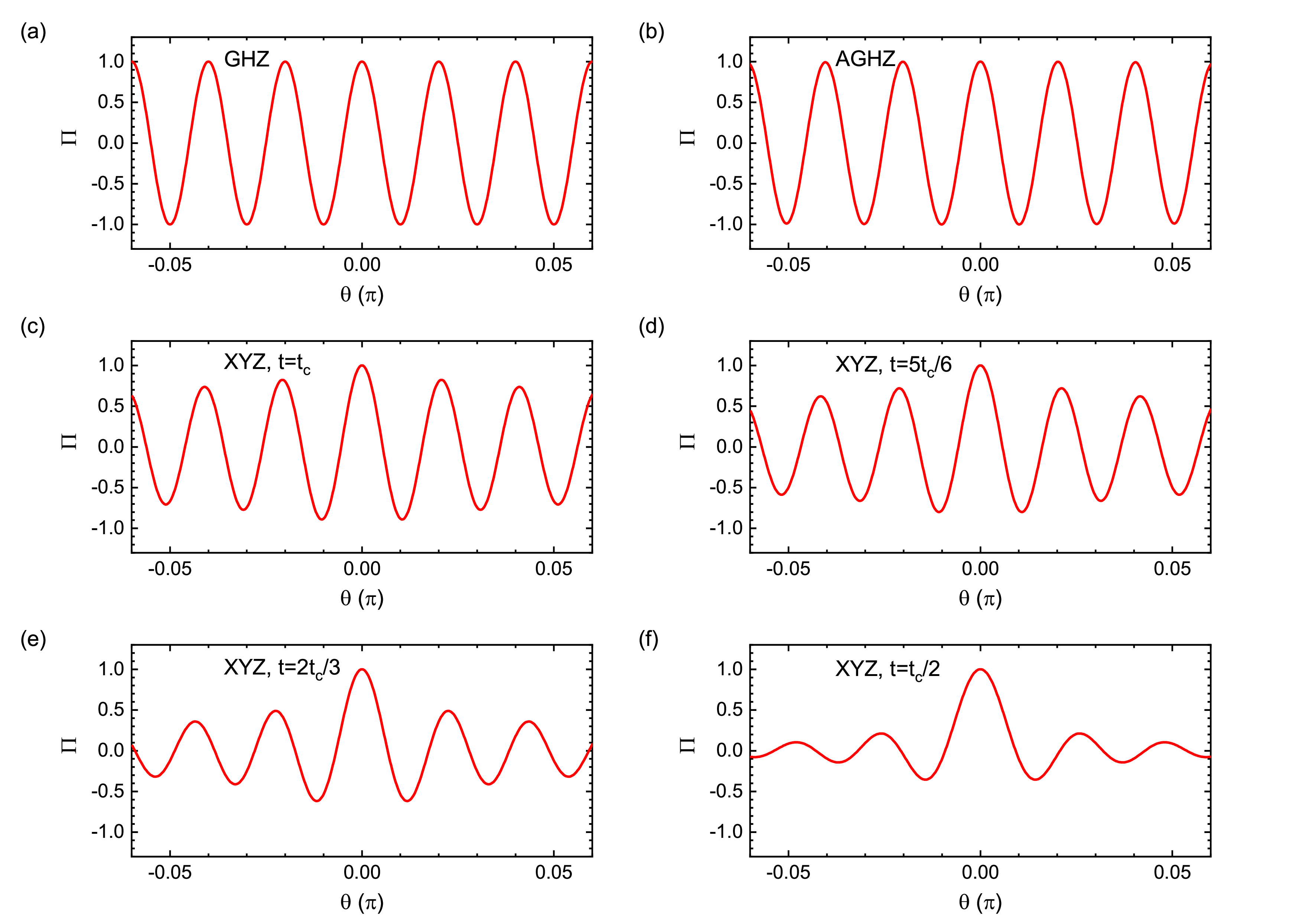}
\caption{Parity oscillation of different GHZ-like state: (a) GHZ state, (b)
approximate GHZ state, (c)-(f) GHZ-like state generated through the XYZ
dynamics for (c) $t=t_c$, (d) $t=5t_c/6$, (e) $t=2t_c/3$ and (f) $t=t_c/2$.}
\label{fig:S3}
\end{figure}

We plot the parity oscillation of different GHZ-like state in Fig.~\ref%
{fig:S3}. The oscillation amplitude of GHZ state remains constant when $%
\theta$ varies, while that of the others decrease as $\abs{\theta}$ gets
larger. For the GHZ-like state generated through the XYZ dynamics, a shorter
evolution time degrades the oscillation amplitude more quickly. In spite of
this, the oscillation of all states in Fig.~\ref{fig:S3} is clear in at
least the first period, providing a possiblity to be applied in
high-precision measurements.

We specifically calculate the angular sensitivity
\begin{equation}
\Delta\theta=\abs{\frac{\Delta\Pi(\theta)}{\partial_\theta\expval{\Pi(%
\theta)}}},
\end{equation}
where $\Delta\Pi(\theta)=(\expval{\Pi^2(\theta)}-\expval{\Pi(\theta)}%
^2)^{1/2} = (1-\expval{\Pi(\theta)}^2)^{1/2}$ and the partial derivative
\begin{equation}
\abs{\partial_\theta\expval{\Pi(\theta)}}=\sum_{m>0}^{N/2}4d_m^2m\sin(2m%
\theta).
\end{equation}
At $\theta=0$, both $\Delta\Pi(\theta)$ and $\partial_\theta%
\expval{\Pi(\theta)}$ vanish, the angular sensitivity can be derived by
taking the limit
\begin{eqnarray}
\lim_{\theta\to0}\abs{\frac{\Delta\Pi(\theta)}{\partial_\theta\expval{\Pi(%
\theta)}}} &=& \lim_{\theta\to0} \frac{\left\{1-\left[d_0^2+%
\sum_{m>0}^{N/2}2d_m^2\cos(2m\theta)\right]^2\right\}^{1/2}}{%
\sum_{m>0}^{N/2}4d_m^2m\sin(2m\theta)}  \nonumber \\
&=&\lim_{\theta\to0}\frac{\left(\sum_{m>0}^{N/2}8d_m^2m^2\theta^2%
\right)^{1/2}}{\sum_{m>0}^{N/2}8d_m^2m^2\theta}  \nonumber \\
&=& \left(4\sum_{m>0}^{N/2}2d_m^2m^2\right)^{-1/2}.
\end{eqnarray}
Since $(\Delta J_z)^2=\sum_{m>0}^{N/2}2d_m^2m^2$, we have
\begin{equation}
\Delta\theta\vert_{\theta\to0} = \left[4\left(\Delta J_z\right)^2\right]%
^{-1/2} = F_Q^{-1/2},
\end{equation}
which is exactly the quantum Cram\'{e}r-Rao bound.

The derivation above indicates that the parity measurement can always be the
optimal measurement for the GHZ-like state generated through the XYZ
dynamics. However, the signal $\partial_\theta\expval{\Pi(\theta)}$ vanishes
at $\theta=0$, thus is impossible to be applied in practice. Instead, we
consider the sensitivity at $\theta_0=\pi/(2N)$, where the signal reaches
its maximum for the GHZ state. The result in Fig.~3(b) shows the quantum Cram%
\'{e}r-Rao bound can still be nearly saturated at this point.

\section{Comparison with the perfect GHZ state}

\begin{figure}[tbp]
    \includegraphics[width=1.0\columnwidth]{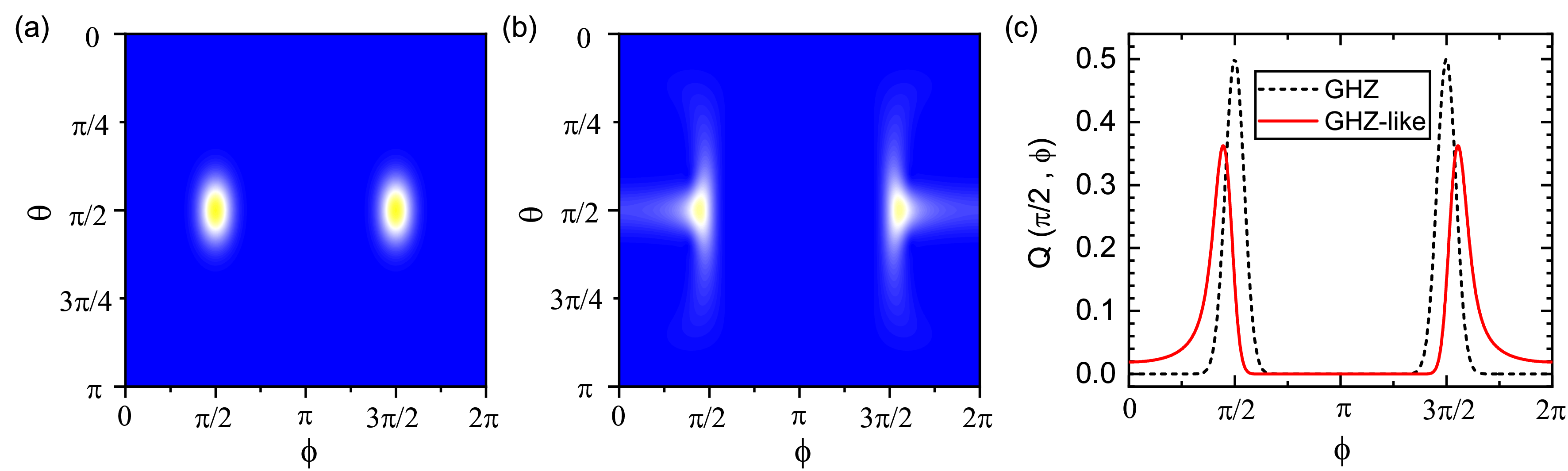}
    \caption{ Husimi $\mathcal{Q}$ function $Q(\theta,\phi)=\abs{\langle \theta,\phi | \psi \rangle}^2$ of (a) the perfect GHZ state and (b) the GHZ-like state, represented with polar coordinates $(\theta,\phi)$. (c) Husimi $\mathcal{Q}$ function along the equator $Q(\pi/2,\phi)$. The perfect GHZ state has been rotated about $x$ axis for $\pi/2$ before calculation.}
    \label{fig:S6}
\end{figure}

In the main text we have shown the GHZ-like state generated using XYZ model shares similar metrological properties with the perfect GHZ state. Here we provide more results for comparison.

The Husimi $\mathcal{Q}$ function can be seen as a quasi-probability distribution of the collective spin. In Fig. \ref{fig:S6} we compare the $\mathcal{Q}$ function of the perfect GHZ state and the GHZ-like state. Both of them concentrate at two opposite poles $(0,\pm1,0)$ on the generalized Bloch sphere, despite the GHZ-like state has an extended uncertainty patch. When focusing on the distribution along the equator, they both exhibit a clear double peak, see Fig. \ref{fig:S6}(c). The peak of the GHZ-like state slightly deviates from $\pi/2,3\pi/2$, and the peak value is lower. Nevertheless, it still shows that the GHZ-like state can be seen as a coherent superposition of two macroscopically distinguishable parts (nearly all spins up and all spins down, if we redefine the axes), which is a significant feature of the perfect GHZ state.

The quantum Fisher information of the perfect GHZ state reaches the maximal value $F_\mathrm{Q}^\mathrm{GHZ}=N^2$, and the GHZ-like state shares the same power law $F^\mathrm{GHZ\text{-}like}_\mathrm{Q}\propto N^2$. In Fig. \ref{fig:S5}(a) we plot the ratio $F_\mathrm{Q}^\mathrm{GHZ\text{-}like}/N^2$ as a function of particle number $N$. We find this ratio increases monotonously as $N$ becomes larger, showing a better match for large particle number. We also analysis the variance $(\Delta J_n)^2$ in directions perpendicular to the optimal rotation axis in Fig. \ref{fig:S5}(b). For the GHZ state $\mathrm{GHZ}=(\ket{N/2}+\ket{-N/2})/\sqrt{2}$, we have $(\Delta J_x)^2=(\Delta J_y)^2=N/4$, as labeled with the black solid curve. It can be found both $(\Delta J_x)^2$ and $(\Delta J_y)^2$ of the GHZ-like state surpass this value, and exhibits a different power law, that is, $(\Delta J_x)^2\sim N^{1.7}$ and $(\Delta J_y)^2\sim N^{1.2}$ given by numerical fitting for $10^2<N<10^3$. This can be seen in its $\mathcal{Q}$ function, which possesses an irregularly shaped uncertainty patch stretching slightly along two orthogonal directions (see Fig. \ref{fig:S6}(b)), rather than an isotropic one.

\begin{figure}
\includegraphics[width=0.9\columnwidth]{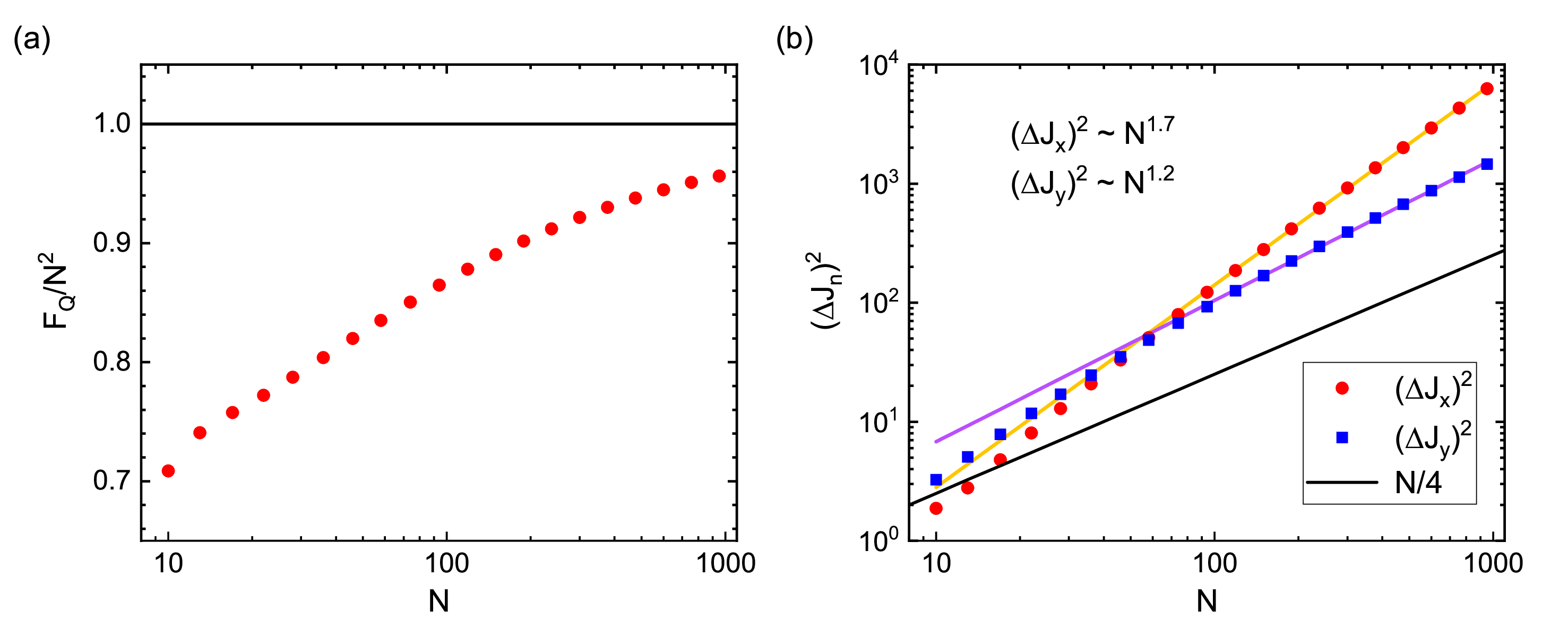}
\caption{Scaling analysis of the GHZ-like state. (a) The ratio $F_\mathrm{Q}/N^2$ as a function of particle number $N$ (red circles). (b) The variance $(\Delta J_n)^2$ in two orthogonal directions perpendicular to the optimal rotation axis. Numerical fitting of $(\Delta J_x)^2$ and $(\Delta J_y)^2$ for $10^2<N<10^3$ is shown with yellow and violet curves, respectively. The result of the perfect GHZ state is labeled with solid black curves. The GHZ-like state has been rotated about $x$ axis for $\pi/2$ before calculation. }
\label{fig:S5}
\end{figure}

The difference between these two states largely results from the fact that there is a certain proportion of probability distributes on $m\neq \pm N/2$ for the GHZ-like state (especially on $m=\pm(N/2-1)$). To see this major distinction, first we prove parity $\Pi$ is a constant in the XYZ dynamics, that is, $\comm{H_\mathrm{XYZ}}{\Pi}=0$. The XYZ Hamiltonian can be expressed in Dicke basis $\{\ket{m}\}$ as
\begin{eqnarray}
    H_\mathrm{XYZ} &=& \frac{2\chi_\mathrm{xyz}}{N}(J_xJ_yJ_z+J_zJ_yJ_x)\\
    &=& \frac{\chi_\mathrm{xyz}}{2iN}\sum_{m}\left[ A^m_j \ket{m+2}\bra{m}+B^m_j \ket{m-2}\bra{m}-2m^2\ket{m}\bra{m}\right] + \mathrm{H. c.},
\end{eqnarray}
where $A_j^m = m\sqrt{(j-m)(j-m-1)(j+m+1)(j+m+2)}$, $B_j^m = m\sqrt{(j+m)(j+m-1)(j-m+1)(j-m+2)}$ and H.c. stands for the Hermitian conjugate. We find
\begin{eqnarray}
    \comm{\ket{m+2}\bra{m}}{\Pi}&=&\sum_n (-1)^n\comm{\ket{m+2}\bra{m}}{\ket{n}\bra{n}}=(-1)^{m}\ket{m+2}\bra{m}-(-1)^{m+2}\ket{m+2}\bra{m}=0,\\
    \comm{\ket{m-2}\bra{m}}{\Pi}&=&\sum_n (-1)^n\comm{\ket{m-2}\bra{m}}{\ket{n}\bra{n}}=(-1)^{m}\ket{m-2}\bra{m}-(-1)^{m-2}\ket{m-2}\bra{m}=0,\\
    \comm{\ket{m}\bra{m}}{\Pi}&=&\sum_n (-1)^n\comm{\ket{m}\bra{m}}{\ket{n}\bra{n}}=(-1)^{m}\ket{m}\bra{m}-(-1)^{m}\ket{m}\bra{m}=0,
\end{eqnarray}
which means every single term of $H_\mathrm{XYZ}$ commutes with $\Pi$, thus we have $\comm{H_\mathrm{XYZ}}{\Pi}=0$.

The fact that $H_\mathrm{XYZ}$ commutes with $\Pi$ forbids the transition between odd and even $m$. Since the initial coherent spin state distributes continually on the basis $\{\ket{m}\}$, we expect about a half of the probability distributes on even $m$ and another half on odd ones. As a result, the probability to find a the GHZ-like state at $m=\pm N/2$ is at most $50\%$, with most of the residue at $m=\pm (N/2-1)$.

However, we can see such a distinction plays no crucial role in its metrological applications, as shown in Fig. 3 in the main text. This is not surprising, since the phase difference of $\ket{\pm N/2}$ accumulated by $e^{iJ_z\phi}$ is $\Delta \varphi=N\phi$, while for $\ket{\pm (N/2-1)}$ it is $\Delta \varphi'=(N-2)\phi$, nearly equal for large $N$.

\section{Comparison with twist-and-turn dynamics}

\begin{figure}[tbp]
\includegraphics[width=0.9\columnwidth]{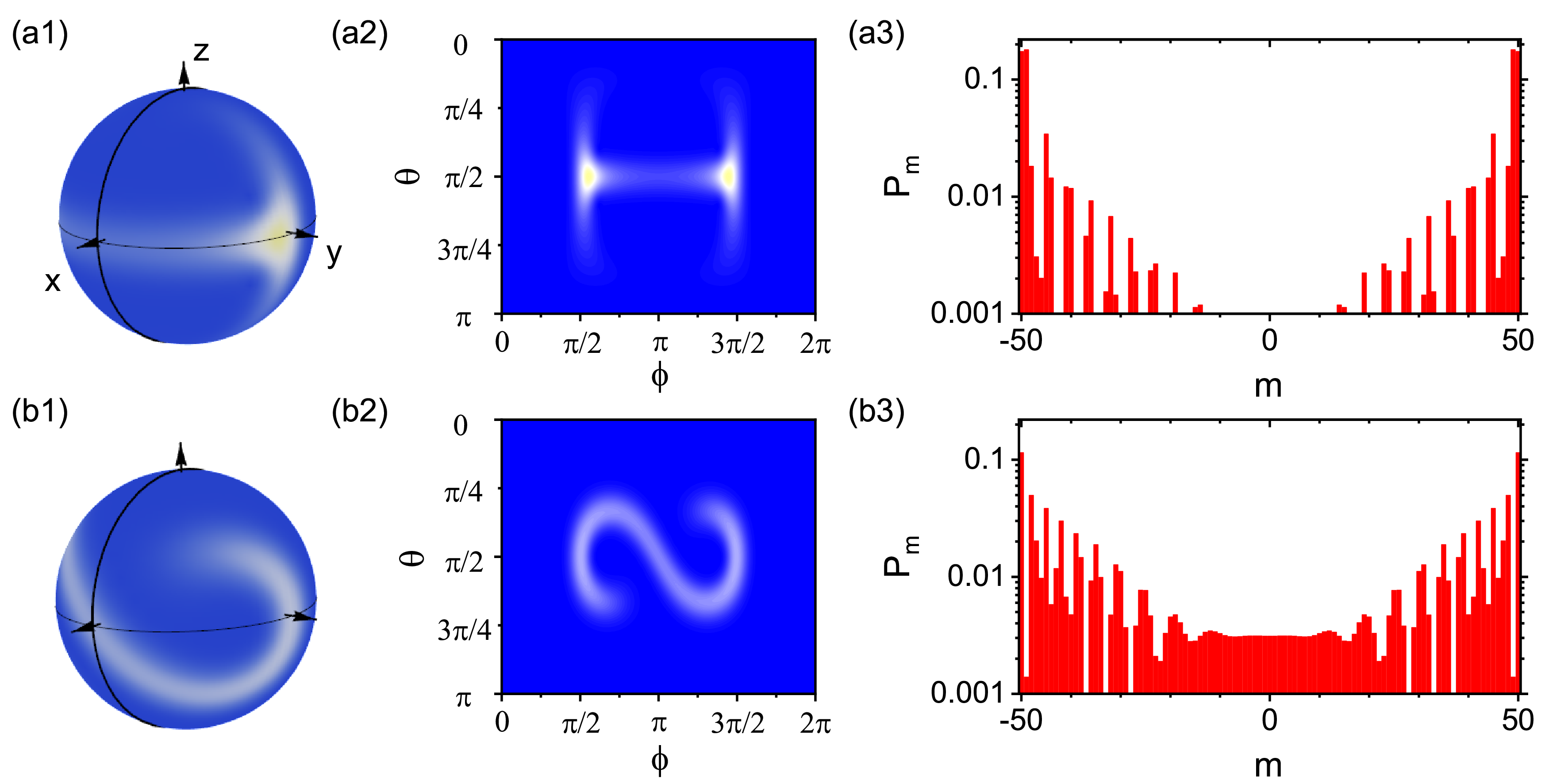}
\caption{Comparison of GHZ-like states generated via (a) XYZ dynamics and
(b) twist-and-turn dynamics for $N=100$ particles. The Husimi $\mathcal{Q}$
function represented (a1 and b1) on the generalized Bloch sphere and (a2 and
b2) with the polar coordinates $(\protect\theta,\protect\phi)$. (a3 and b3)
The probability distribution on eigenstates of $J_y$.}
\label{fig:S4}
\end{figure}

As proposed in Ref.~\cite{Micheli2003}, GHZ-like states can be created via
twist-and-turn (TNT) dynamics
\begin{equation}
H_\mathrm{TNT}=\chi J_y^2 + \Omega J_x,
\end{equation}
which is realized by adding a constant turning term $\propto J_x$ to an
original OAT Hamiltonian. The optimal performance appears when $\Omega=\chi
N/2$ and it needs $\chi t\simeq\ln(8N)/N$ to obtain the GHZ-like state. 
For large $N$, the evolution time of the TNT dynamics scales as $\propto \ln(N)/N$, which is the same as our scheme.

To make a comparison between GHZ-like states generated using these two
models, in Fig.~\ref{fig:S4} we plot Husimi $\mathcal{Q}$ functions and
probability distributions of them. It can be easily found that the XYZ model
creates GHZ-like state which is much more concentrated at two poles $%
(0,\pm1,0)$ on the generalized Bloch sphere than the TNT scheme. Another
major distinction appears at the probability distribution near $m=0$, where
that of the XYZ model nearly vanishes, while a certain degree of probability
still exists for the TNT model. As a result, the two parts of the GHZ-like
state seems fully separated for the XYZ model but is connected to some
extent for the TNT model. We can therefore conclude that the XYZ scheme
produces GHZ-like state which is much closer to the perfect GHZ state in a
same time scale with the TNT scheme.

\end{widetext}

\end{document}